# Consequences of catastrophic disturbances on population persistence and adaptations

Simone Vincenzi[a], Michele Bellingeri[a]

[a]Dipartimento di Scienze Ambientali, Università degli Studi di Parma, Viale G. P. Usberti 33/A, I-43100 Parma, Italy.

**Corresponding author:**
Simone Vincenzi,
Dipartimento di Scienze Ambientali
Università degli Studi di Parma
Viale G.P. Usberti 33/A
I-43125 Parma
Tel.: +39 0521 905696
Fax.: +39 0521 906611
email: simon.vincenz@gmail.com

## Abstract

The intensification and increased frequency of weather extremes is emerging as one of the most important aspects of climate change. We use Monte Carlo simulation to understand and predict the consequences of variations in trends (i.e., directional change) and stochasticity (i.e., increase in variance) of climate variables and consequent selection pressure by using simple models of population dynamics. Higher variance of climate variables increases the probability of weather extremes and consequent catastrophic disturbances. Parameters of the model are selection pressure, mutation, directional and stochastic variation of the environment. We follow the population dynamics and the distribution of a trait that describes the adaptation of the individual to the optimum phenotype defined by the environmental conditions

The survival chances of a population depend quite strongly on the selection pressure and decrease with increasing variance of the climate variable. In general, the system is able to track the directional component of the optimum phenotype. Intermediate levels of mutation generally increase the probability of tracking the changing optimum and thus decrease the risk of extinction of a population. With high mutation, the higher probability of maladaptation decreases the survival chances of the populations, even with high variability of the optimum phenotype.

## 1. Introduction

With climate change, many species will experience selection pressures in new directions and at new intensities, and the degree to which species respond adaptively will have an important influence on their capacity to survive over the coming decades and millennia [1].

Changes in the long-term mean state of climate variables (i.e, climate trends) and their consequences on survival, evolution and adaptation of species have been intensively studied for more than 20 years [2]. The intensification of weather extremes is emerging as one of the most important aspects of climate change [3] and the debate is expanding from an analysis of trends to an interest in extreme events and associated catastrophic disturbances, such as periods of heavy rainfall (with associated floods and landslides), fires, droughts, heat waves [4]. Catastrophes are characterized by statistical extremity, timing, and abruptness relative to the life cycles of the organisms affected; they can disrupt ecosystems, communities, or population structure and change resource pools, substrate availability, or physical environment [3, 5, 6]. Many adaptations (in life histories, morphological or behavioral traits) can be associated with catastrophic disturbance events [7].

There is increasing evidence that the frequency and severity of climate extremes and associated catastrophes have already increased in several regions [8, 9]. Hence, there is the urgent research need to meet the challenges posed by extreme events and catastrophic disturbances. Despite this, their evolutionary consequences have largely been unexplored (e.g., [10]), and more attention has been paid by ecologists to exploring adaptations of individuals to changing trends in climate variables (e.g., [11]), abrupt changes in the environment [12] or smooth and periodic changes [13,14].

Recent individual-based quantitative genetic models with stochastic dynamics [15] suggest that evolution may quickly rescue populations after they collapse under abrupt environmental change. Fitness should initially decline after the environmental change, but then recover through adaptation [16]. According to theory, whether populations can be rescued by evolution depends upon several crucial variables, including population size, genetic variation within the population, and the degree of maladaptation to the new environment [17]. Using Monte Carlo simulations, Bena et al. [14] compared the effects of a smooth variation of the optimum phenotype (as determined by environmental conditions) with those emerging from an abrupt change of the environment and found that sufficiently large mutation rates can increase substantially the probability of population persistence in both scenarios.

In many systems, selection is not only directional (e.g., higher temperatures, higher rainfall) but fluctuates (e.g., through cycles, stochastically). Climate change models show that the variance of climate variables like temperature or rainfall may change much more dramatically than their means [18] and will thus intensify the stochastic component of selection. Shifts and excursions might cause some populations to perpetually chase (evolutionarily) alternate optimal phenotypic extremes. Such populations would face a demographic cost if evolution during one environmental phase resulted in maladaptation and reduced favorable genetic variation with respect to the next. This could be exacerbated by strong selection pressure and little opportunity for the emergence of new phenotypes (i.e., high hereditability of traits and low mutation rates). With climate change, it is possible that by chance the population will experience a long sequence of particularly extreme environments [3]. This may cause the population growth rate to be negative for a long enough time to cause extinction. In addition, even when extinction does not follow immediately the extreme event, the loss of genetic variability resulting from a low population size can substantially reduce the population's ability to respond to future selective challenges and increases the chances of an extinction vortex [19].

Despite the inherent difficulties of predicting the impact of climate change on species persistence and evolutionary trajectories, the general trends and dynamics at the individual- and population-level are reasonably comprehensible and modeling can provide probabilistic expectations for population dynamics and evolutionary processes. Here, we want to investigate the implications of selection pressure and mutation rates on the behavior of simulated populations living in a habitat subject to changing selective processes. Monte Carlo simulation represents a valid tool to understand and predict the consequences of variations in trends (i.e., directional change) and stochasticity (i.e., increase in variance) of climate variables and consequent selective processes by using simple models of population dynamics.

## 2. Model of population dynamics

We consider a population that consists of hermaphrodite individuals living in a spatially-extended habitat modeled as a vector of length *K*, where *K* is the carrying capacity of the system (i.e., maximum number of individuals supported). This means that only one individual can occupy each element $j = 1 \ldots K$ of the vector, and introduces density-dependent population regulation through a ceiling effect, as described below. We assume that individuals cannot move, therefore an individual occupies the same element during the simulation.

The populations has discrete generations (i.e., reproduction is discrete in time) and is composed of N($t$) individuals. Generations are overlapping, meaning that parents do not die after reproducing. Each individual is characterized by a single quantitative trait $\phi$ with value ranging from 0 to 1. The population lives in a habitat characterized by an optimum phenotype $\Theta(t)$ that exhibits temporal change. This is assumed to result from variations in a climate variable, such as rainfall or temperature, selecting for a phenotype. The degree of maladaptation between the optimum phenotype $\Theta(t)$ and a single trait $\phi_i$ defines the fitness of an individual. The time step is one year.

In general, the temporal change of the optimum phenotype may be either directional, stochastic or a combination of both. A simple model for this is a optimum phenotype $\Theta(t)$ that moves at a constant rate $\beta_\mu$ per year, fluctuating randomly about its expected value $\mu(t)$. We thus introduce a directional and stochastic temporal change of the optimum phenotype (Fig.1a). $\Theta(t)$ is randomly drawn at each time step from a normal distribution $N(\mu(t), \mathrm{sd}(t))$, where $\mu(\mathrm{t}) = \mu_0 + \beta_\mu t$ and $sd(t) = sd_0 + \beta_{\mathrm{sd}} t$.

In the context of climate variables, extreme weather events causing catastrophic disturbances (i.e., very large deviations of a system's behavior from the habitual one) in a sequence of independently and identically distributed random variables are either the maximal values in a time window or they are defined by overcoming a predefined threshold (threshold crossing) [20]. In our model, values of $\Theta(t)$ outside (0,1) represent an extreme event causing a catastrophic disturbance to which the vast majority of individuals cannot be adapted, thus causing a population collapse (i.e., strong reductions in population size). This may be interpreted as a catastrophic flood following an exceptional rainfall or a heat wave caused by high temperatures.

Our model is similar in spirit to the one used by Droz and Pekalsky [21]. The fitness of an individual *i* with trait value $\phi_i$ is:

$$f(i) = |\phi_i - \Theta(t)| \qquad (1)$$

The probability *p* of an individual *i* with fitness f(*i*) survive to next year is:

$$p(i) = \exp(-(s * f(i))^2) \qquad (2)$$

where *s* is the selection pressure. With increasing *s* the habitat is more demanding (for a given fitness *f* the probability of survival decreases). Since no individual can be perfectly adapted to the moving optimum phenotype $\Theta$, we did not account for a decrease in survival probability with age (in case of constant $\Theta$ over simulation time, accounting for it would be necessary to avoid the presence of individuals living forever).

Offspring inherit the trait $\phi$ from its parents $p_1$ and $p_2$ as follows:

$$\phi_0 = 0.5\ (\phi_{p1} + \phi_{p2}) + M\varepsilon \qquad (3)$$

where $\phi_0$ is the trait value of the offspring, $\phi_{p1}$ and $\phi_{p2}$ are the trait values of the parents, $M$ represents mutation-segregation-recombination [22] and $\varepsilon$ is random number drawn from a uniform distribution bounded between (-1,1). We will refer to $M$ as simply mutation.

The Monte Carlo simulation at a time $t$ during the simulation proceeds as follows:

1) We draw the optimum phenotype $\Theta(t)$ from $N(\mu(t), \mathrm{sd}(t))$.
2) We compute the fitness of individuals by applying Eq (1) and calculate their survival probability by applying Eq (2).
3) We define the survival of individuals with Bernoulli trials.
4) We compute the total number of individuals alive $N(t)$ and check the distribution of trait $\phi$ in the population. A population is considered extinct if at any time during the simulation there are less than ten individuals left.
5) We pick the first individual alive starting from $j = 1$. When the individual $j$ is alive, we check if the $(j+1)$ individual is alive. If yes, the parents $j$ and $(j+1)$ produce randomly from 1 to 4 offspring (we chose 4 as the maximum number of offspring produced by following a pattern-oriented procedure [23] to allow for a quick rebound of population size after a strong reduction caused by an extreme event). If no, the individual *j* does not reproduce. This introduces the Allee effect [24], that is a positive density-dependent effect at low densities through higher mating opportunities. Then, we proceed to ($j$+2) and repeat the procedure up to $j = K$.

6) As we assume that the optimum phenotype $\Theta(t)$ defines the whole time-step, we applied steps (2) and (3) to offspring. Following an example provided above, a heat wave affects the survival of both adults and offspring. This further intensifies the selective consequences of the optimum phenotype. Offspring are placed randomly on the empty elements of the vector to avoid spatial autocorrelation. When all the empty elements have been occupied, the remaining offspring die (density-dependence through a ceiling effect). Offspring at year $t$ become adults at year $t$+1 and are able to reproduce.

Our simulation model has the following control parameters: carrying capacity of the environment $K$, mutation $M$, selection pressure $s$ and the parameters which govern directional and stochastic variations of the optimum phenotype, that is $\mu_0$, $\beta_\mu$ , $sd_0$ and $\beta_{sd}$. To simplify the interpretation of results, we set some of the parameters. For each replicate: $K = 2000$, $\mu_0 = 0.5$, $\beta_\mu = 0.001$ and $sd_0 = 0.1$. Simulations were performed for combinations of $s$ (from 2.5 to 3.5), $M$ (from 0 to 0.2) and $\beta_{sd}$ taking values 0.0005, 0.0010, 0.0015, 0.0020 (scenarios of increasing variability of $\Theta$ over simulation time, Fig. 1b).

Every simulation replicate lasts 100 years and starts with 500 individuals with trait $\phi$ drawn from a uniform distribution bounded between (0,1) (Fig.2). We chose a population size of 500 individuals (one fourth of carrying capacity) because we wanted to explore the space of parameters allowing for extinctions in the first years of simulation.

We use different quantities to characterize the behavior of the simulated populations. At the level of single replicates, we recorded:

i) Extinction or survival (coded as a binary variable equal to 0 for persistence and 1 for extinction).

ii) Time of extinction, obviously recorded only for the populations going extinct during simulation time.

iii) Time-dependent value of the trait $\phi$, and in particular the mean value of $\phi$ (ranging from 0 to 1) at the end of simulation time, only when the population did not go extinct.

We did not focus on the number of individuals at the end of simulation time (100 years) since it was largely determined by the succession of $\Theta$ near the end of simulations (Fig. 2).

For an ensemble of realizations (100 replicates for a fixed set of parameters), we computed:

a) Frequency of population extinction, computed as the number of replicates in which the population went below ten individuals during simulation time.

b) Mean time to extinction (for the populations which went extinct during simulation time).

c) Mean across replicates of the mean value of trait at the end of simulation time, for the replicates in which the populations did not go extinct.

## 3. Results and discussion

In Fig. 1b we show the probability of catastrophes with the different scenarios of variability of $\Theta$. The probability of a catastrophe, that is of optimum phenotype $\Theta(t)$ outside (0,1), reaches a maximum of 0.12 at the end of simulation time ($t$ = 100) for the most variable scenario ($\beta_{sd}$ = 0.0020). With the parameters we chose, there is a higher probability of extreme events in the same direction as directional change (more values of $\Theta > 1$ than $< 0$ are expected), although the probability of both events increases over the simulation time (Fig. 1a). In other words, with increasing temperatures there is a higher probability of heat waves than of cold waves and with increasing rainfall (and thus increasing flows) there is an higher probability of floods than of droughts.

The consequences of different values of $\beta_{sd}$ for the probability of extreme events is clear after the first 40-50 years of simulation while little difference among scenarios in the probability of extreme events can be noted before that time. In Fig. 2 we report examples of replicates for the four scenarios of variability. For all replicates we set $s$ = 3 and $M$ = 0.1 (thus intermediate values for both parameters). With higher values of $\beta_{sd}$ the population shows repeated collapses. Selection tries to bring the average trait close to the instantaneous optimum, while mutation introduces diversity and broadens the distribution of the trait. There is a clear shift in all replicates of the mean value of trait $\phi$ toward 0.6 over simulation time - which is the value taken by $\mu$ when $t$ = 100 - even in high variability scenarios. The only exception is the scenario with the highest variability, in which at $t$ = 100 there is a mean value of trait $\phi$ in the proximity of 0.5. In the specific example provided, a few years of $\Theta$ below 0.6 "push back" the trait $\phi$ toward lower values than those selected for by the directional component of $\Theta$.

As noted by Siepielski et al. [25], the "temporal landscape" of selection across taxa shows that the strength and the direction of selection often vary through time, even in absence of climate change. Especially with strong selection pressure and high variability of the optimum $\Theta$, alternating selection over time might cancel out periods of directional selection such that effective selective (quasi) neutrality of trait variation is maintained over time. However, after a single extreme event or a succession of them, this balancing effect does not occur, leading to directional changes in trait frequency within the population. Apart from the contribution of directional change, the distribution of trait $\phi$ is "pulled" toward higher values over simulation time, since there is a higher probability of extreme events in the same direction as directional change (as previously discussed).

In Fig. 3 we present a phase diagram of equal probability of extinction in the mutation-selection plane for each scenario of variability of $\Theta$. The survival chances of a population depend quite strongly on the selection pressure and decrease substantially with increasing $\beta_{sd}$ for the same selection-mutation combinations, indicating that populations could rarely adapt to a strong linear increase in variance of $\Theta$.

There is a range of the selection pressure values within all scenarios of variability in which populations have some probability to persist (Fig. 3). Outside this range, broadly for $s$ higher than 2.8, the probability of extinction increases in all scenarios. If selection is too strong, then the distance between the average phenotype and the optimum is small at any time during simulation, but the decrease in population size induced by selection may be too high for population persistence. If the selection is weaker, fewer individuals die from ill-adaptation and the population can persist with a greater diversity in trait $\phi$.

For $\beta_{sd}$ = 0.0020 populations can survive only with intermediate levels of mutation and very low selection pressure, while extinction is the inevitable outcome for all other selection-mutation combinations (Fig. 3). The adaptive value of intermediate levels of mutation is clear also for $\beta_{sd}$ = 0.0005 and for $\beta_{sd}$ = 0.0010, while for $\beta_{sd}$ = 0.0015 the only clear pattern is along a selection gradient. It appears from Fig. 3 that increasing mutation amplitude is adaptive up to intermediate values, while higher mutation values are not adaptive (they increase the probability of population extinction).

Contrary to our results, Bena et al. [14] found that mutation is unfavorable to the survival of a population in a constant environment, since it increases the probability of a mismatch of offspring phenotype to the environment optimum, even though the parents might be well-adapted. Therefore, any level of mutation will result in the production of non-optimal trait in a constant environment (given an adapted population), but it will increase the probability of tracking a moving optimum and thus increase the survival chances of a population. According to our results, even in presence of high variability of the optimum phenotype $\Theta$, high mutation increases the probability of losing adaptations in the next generation and thus decreases the probability of population persistence. When mutation is low, the population cannot track the variations of $\Theta$. In conclusion, for both mutation extremes (high or low mutation) there is an increase in the probability of maladaptation, albeit for different reasons, and consequent risk of extinction.

The influence of selection, mutation and $\beta_{sd}$ on the average time to extinction is reported in Fig. 4. For $\beta_{sd}$ = 0.0005, for the few populations going extinct with intermediate mutations, this happens only in the first years of simulation after an unfavorable succession of alternate $\Theta$ (direction of selection varying through time). With intermediate selection pressure, populations go extinct mostly at the end of simulation time, when an increase of occurrence of extreme values is expected for all scenarios of variability (Fig. 1). An increase in selection pressure tends to decrease time of extinction in all scenarios of variability.

In general, the system is able to track the directional component of the optimum (Fig. 5). The mean value of trait $\phi$ at the end of simulation time does not depend on selection, therefore even for very small selective pressure and in presence of sufficient mutation $M$, the mean value of trait $\phi$ follows the directional component of $\Theta$ (Fig. 6). With no mutation or very low mutation, there is little potential for adaptive shifts and thus the mean value of $\phi$ is largely determined by the optimum phenotypes in the first few years (Fig. 6). For $\beta_{sd}$ = 0.0005 and $\beta_{sd}$ = 0.0010 the mean value of trait $\phi$ in the population increases, and thus tracks the changes in $\mu$(t), also for very high mutation. In contrast, for $\beta_{sd}$ = 0.0015 and $\beta_{sd}$ = 0.0020 the mean value of trait $\phi$ increases with increasing mutation, but with very high mutation the mean value of trait $\phi$ tends to be lower than in scenarios with lower variability. Since in an substantial fraction of replicates with high mutation the population went extinct (Fig. 3), we cannot exclude that for only a particular sequence of $\Theta$ near the end of simulation time (resulting in mean value of trait close to 0.5) the populations were able to persist, thus preventing more general insights.

## 4. Conclusions

Extreme events occur in all systems with complex dynamics, but the details of the creation of these large fluctuations are still rarely understood, and therefore their prediction, including that of their consequences on natural populations, remains a challenge. However, many significant impacts of climatic change are likely to come about from shifts in the intensity and frequency of extreme weather events and the prediction of their effects on population dynamics and evolution of traits in natural populations call for wide and intense scientific investigations. These events may result in rapid mortality of individuals and extinction of populations or species [26, 27, 28, 29, 30] and changes in community structure and ecosystem function [31, 32, 33]. Variations in disturbance timing, predictability, frequency and severity make difficult to predict sign and strength of selection [10]. In some cases, catastrophic events may be so swift or severe that there is little possibility for adaptive responses, with population extinction being the inevitable result. However, given sufficient evolutionary potential (i.e., genetic variation within a populations), models suggest that species can survive the effects of extreme events [34]. However, if variability of the optimum phenotype is too high, a relevant potential for extinction exists even when populations might possess genetic variation for adaptation.

Despite simplifying the life-cycle of a natural population, the model we have presented here provides a useful starting point for the investigation of the potential of the populations to adapt (and survive) to an increase in the variability of environmental conditions. The simulations showed that the probability of survival of populations is dramatically affected by slight increases of the variance of the optimum phenotype. Although not universal across scenarios of variability, intermediate mutation seem to be adaptive, while increasing selection pressure consistently decreases the probability of population persistence.

## Acknowledgements

The authors thank Luca Bolzoni and Kate Richerson for discussion and comments which greatly increased the quality of the manuscript.

# Figure Captions

Fig. 1 – *Weather extremes.* (a) Expected increase in the probability of occurrence of extreme weather events with climate change (gray areas) for an hypothetical climate variabile (e.g., rainfall, temperature), as defined in our model. Solid line represent current scenario ($\mu$ = 0.5, sd = 0.1) while dotted line represent a future scenario at the end of simulation time (dotted line, $\mu$ = 0.6, sd = 0.25). Jentsch et al. [3] and Smith [4] provided similar representations. (b) Expected probability of optimum phenotype $\Theta(t)$ outside (0,1) for different changes in variability during simulation time. Solid line - $\beta_{sd}$ = 0.0005; short-dashed line - $\beta_{sd}$ = 0.0010; long-dashed line - $\beta_{sd}$ = 0.0015; dashed-dotted line - $\beta_{sd}$ = 0.0020.

Fig. 2 – *Examples of simulations*. Examples of simulation for the four scenarios of variability with selection pressure $s$ = 3 and mutation $M$ = 0.1. The optimum phenotype $\Theta(t)$ is randomly drawn at each time step from a normal distribution $N$ ($\mu(t)$,sd($t$)), where $\mu(t) = \mu_0 + \beta_\mu t$ and $sd(t) = sd_0 + \beta_{sd} t$. The histograms represent the distribution of trait $\phi$ at $t$ = 1, 20, 40, 60, 80, 100. The vertical dashed line is set at 0.5. The mean value of trait of the population tracks the directional change ($\mu$ = 0.6 at $t$ = 100) in all the examples of simulation except for $\beta_{sd}$ = 0.0020 at $t$ = 100. The fluctuations in population size tend to increase with increasing $\beta_{sd}$, parallel to increase in fluctuations of optimum phenotype $\Theta(t)$.

Fig. 3 – *Phase diagram for extinction probability.* Phase diagram of equal probability of extinction in the mutation-selection plane for the four scenarios of variability of $\Theta$ ($\beta_{sd}$ = 0.0005, 0.0010, 0.0015, 0.0020). The frequency of population extinction for combinations of selection pressure *s* and mutation *M* is computed as the number of replicates in which the population did go below ten individuals during simulation time.

Fig. 4 - *Phase diagram for mean time to extinction.* Phase diagram of equal mean time to extinction in the mutation-selection plane for the four scenarios of variability of $\Theta$ ($\beta_{sd}$ = 0.0005, 0.0010, 0.0015, 0.0020).

Fig. 5 - *Phase diagram for mean value of trait.* Phase diagram of equal mean across replicates of the mean value of trait $\phi$ at the end of simulation time in the mutation-selection plane for the four scenarios of variability of $\Theta$ ($\beta_{sd}$ = 0.0005, 0.0010, 0.0015, 0.0020). The mean was computed only for the populations which persisted up to the end of simulation time. The white region in the phase diagram for $\beta_{sd}$ = 0.0020 indentifies mutation-selection combinations for which populations had no chances to persist up to end of simulation time (see Fig. 3).

Fig. 6 – *Distribution of trait for increasing mutation.* Examples of the distribution of trait $\phi$ for increasing mutation *M* at the end of simulation time. All simulations performed with *s* = 3 and $\beta_{sd}$ = 0.0010.

# Figure 1

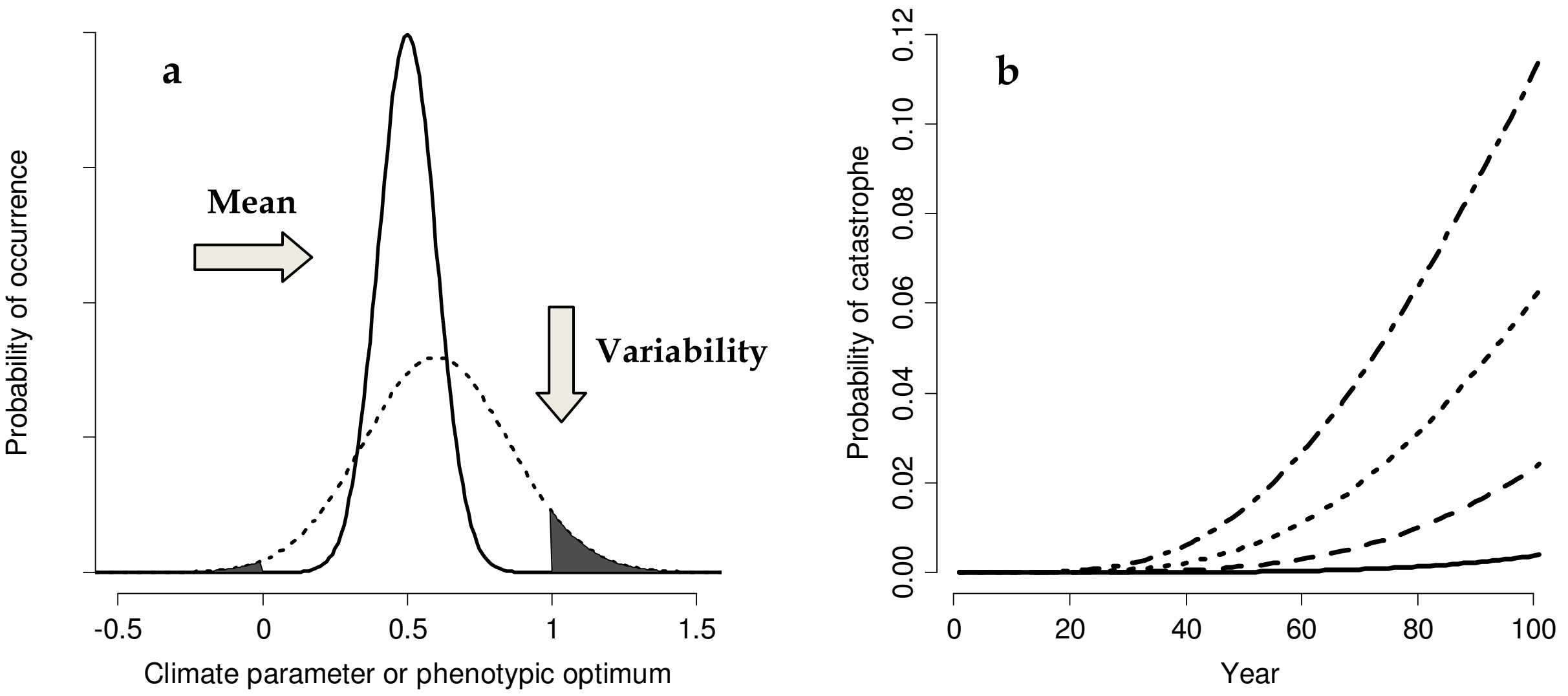

# Figure 2

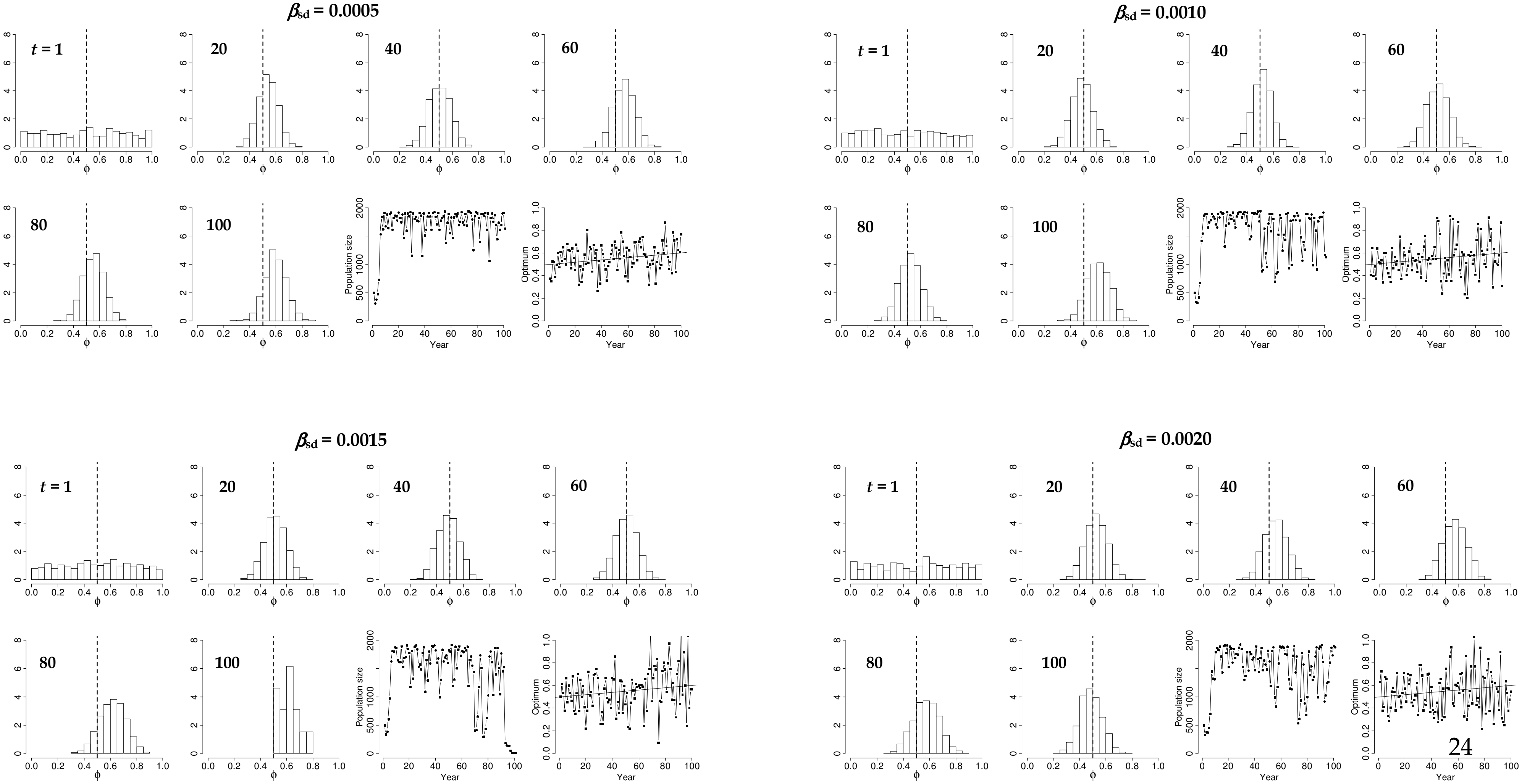

# Figure 3

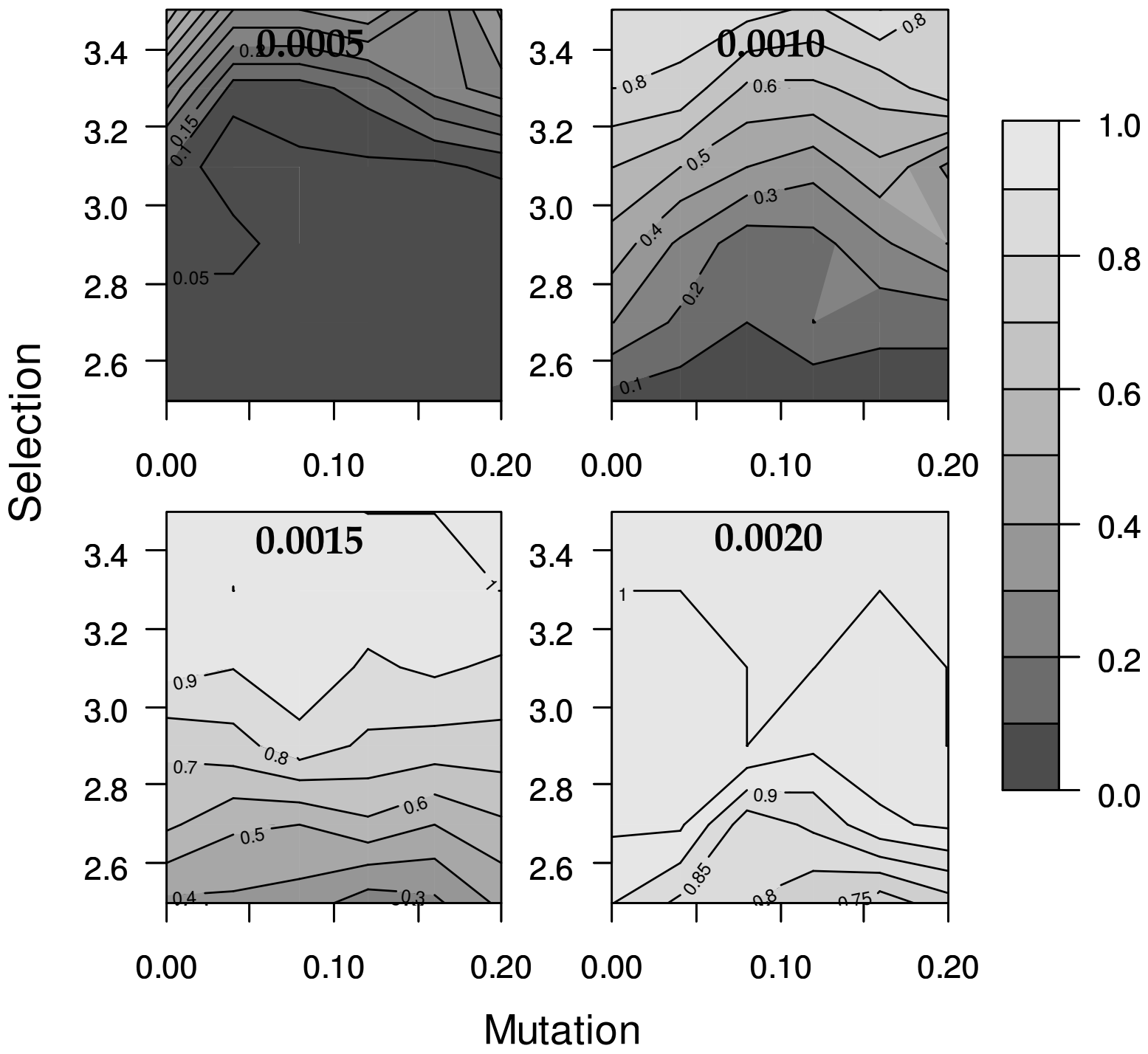

# Figure 4

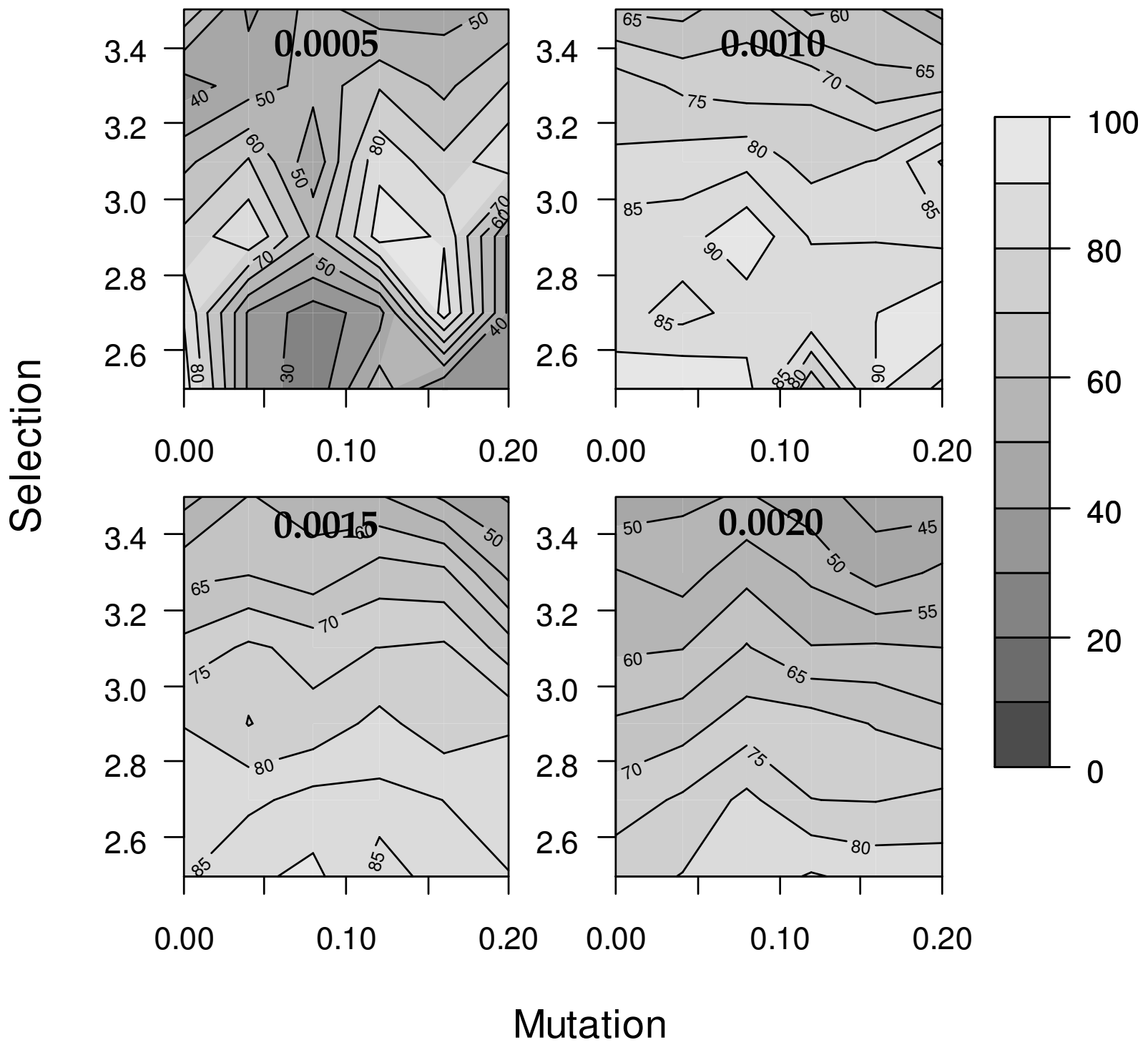

# Figure 5

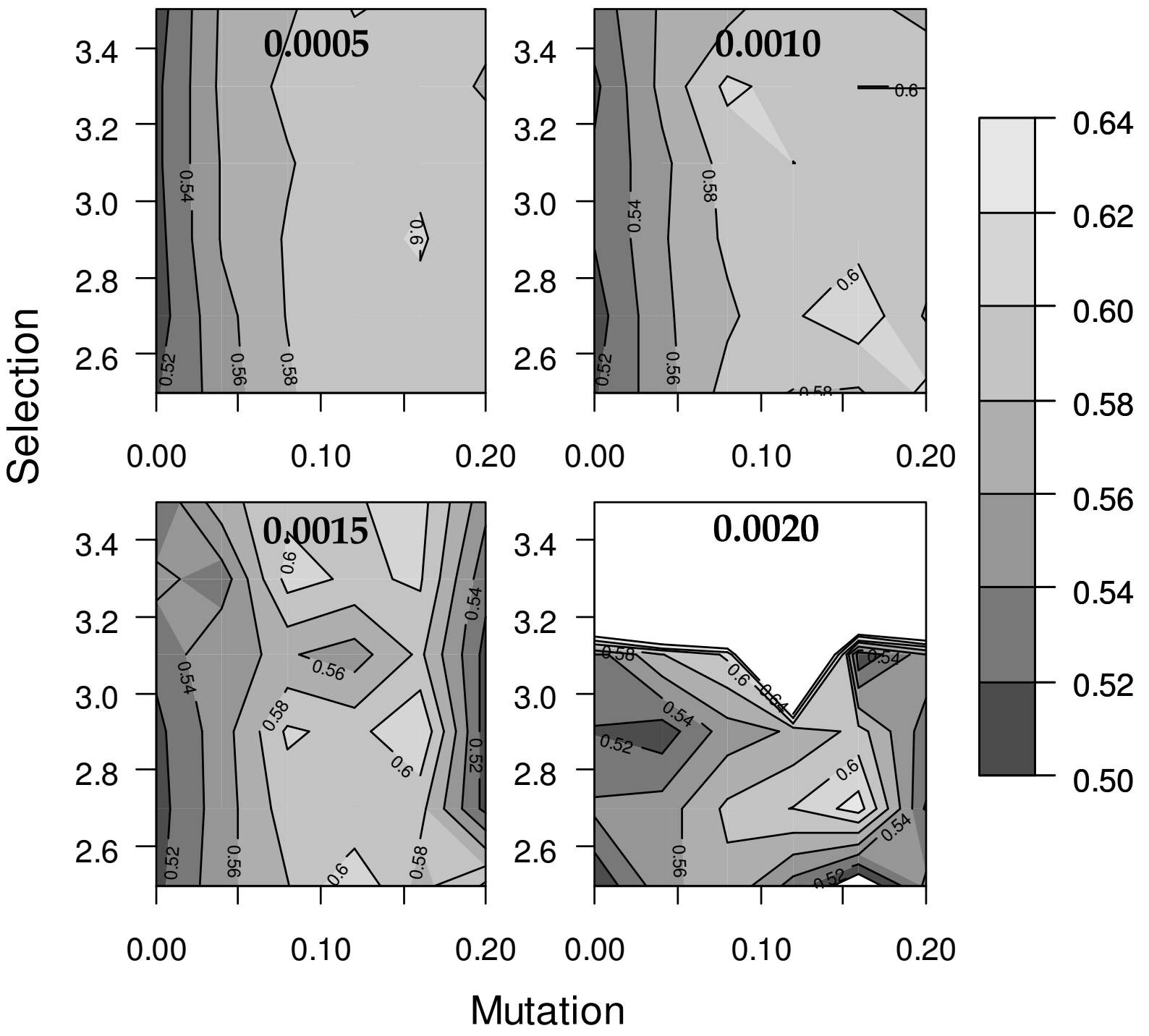

# Figure 6

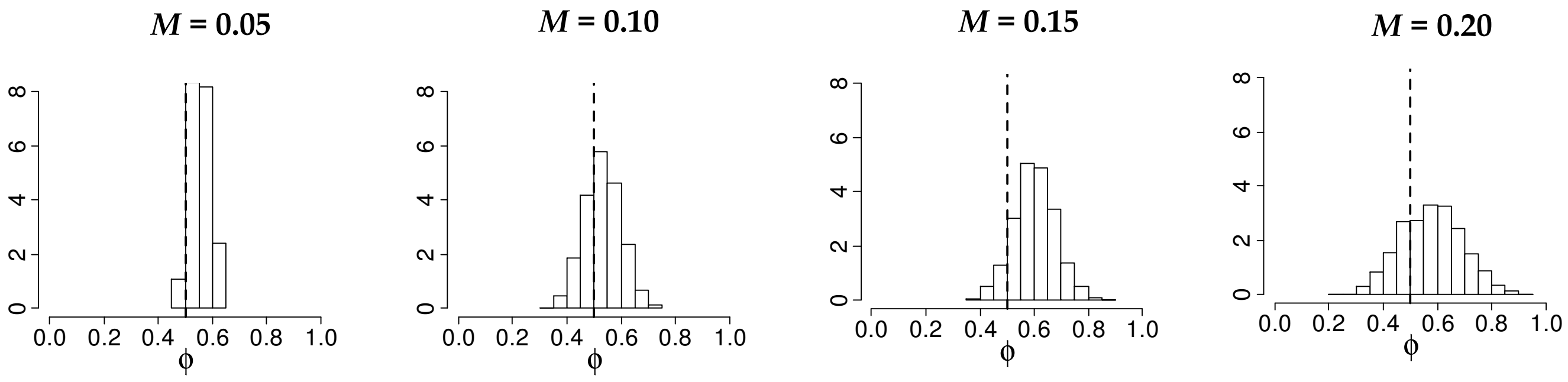